\documentclass[aps,prb,nobibnotes,reprint,twocolumn,superscriptaddress]{revtex4-2}

\usepackage{amsmath, amssymb, amsfonts}
\usepackage{array, booktabs, tabularx}
\usepackage{xcolor, graphicx, tikz}
\usepackage[caption=false]{subfig}
\usepackage{hyperref}  
\usepackage{braket}
\usepackage{bm}
\usepackage[ruled]{algorithm2e}
\usepackage[normalem]{ulem} 

\setcitestyle{super, sort&compress}

\hypersetup{colorlinks = True, allcolors = {blue}}

\newcommand{\FC}[1]{ c_{#1}^\dagger }
\newcommand{\FA}[1]{ c_{#1} }
\newcommand{\FN}[1]{ \hat{n}_{#1} }
\newcommand{\RFC}[1]{ f_{#1}^\dagger }
\newcommand{\RFA}[1]{ f_{#1} }

\newcommand{\FClt}[1]{ a_{#1}^\dagger }
\newcommand{\FAlt}[1]{ a_{#1} }
\newcommand{\FCrt}[1]{ b_{#1}^\dagger }
\newcommand{\FArt}[1]{ b_{#1} }
\newcommand{\FOv}[1]{ S_{#1} }

\newcommand{\FNbar}[1]{ \bar{n}_{#1} }
\newcommand{\FS}[1]{ \phi_{\leftarrow #1} }

\newcommand{\SP}[1]{ S_{#1}^{+} }
\newcommand{\SM}[1]{ S_{#1}^{-} }
\newcommand{\SZ}[1]{ S_{#1}^{z} }

\def\mi{ \mathrm{i} }
\def\EYE{ \mathbb{I} }
\def\HermConj{\text{h.c.}}
\newcommand{\ComCom}[1]{\mathcal{O}(#1)}

\def\su2{$su(2)$}
\def\spinhalf{\text{spin-1/2}~}

\newcommand{\Eq}[1]{Eq.~({#1})} 
\newcommand{\Fig}[1]{Figure~{#1}}

\newcommand{\Sec}[1]{Section~{#1}}
\newcommand{\Reference}[1]{Ref.~{#1}} 
\newcommand{\Appx}[1]{Appendix~{#1}}

\def\PCSD{Physical and Computational Sciences Directorate, Pacific Northwest National Laboratory, Richland, Washington 99354, United States}
\def\UWChem{Department of Chemistry, University of Washington, Seattle, Washington 98195, United States}

\begin{document}


\title{Fermionic mean-field dynamics for spin systems beyond free fermions}

\newcommand{\corrnote}{
Authors to whom correspondence should be addressed:
\href{mailto:rishab.dutta@pnnl.gov}{rishab.dutta@pnnl.gov},
\href{mailto:marc.illasubina@pnnl.gov}{marc.illasubina@pnnl.gov}.
}

\author{Rishab Dutta}
\thanks{\corrnote}
\affiliation{\PCSD}

\author{Marc Illa}
\thanks{\corrnote}
\affiliation{\PCSD}

\author{Niranjan Govind}
\affiliation{\PCSD}
\affiliation{\UWChem}

\author{Karol Kowalski}
\affiliation{\PCSD}


\begin{abstract}

We introduce the fermionized time-dependent Hartree--Fock (fTDHF), a real-time quantum dynamics method for \spinhalf Hamiltonians following their mapping to fermions via the Jordan--Wigner transformation. 
fTDHF is formally equivalent to exact dynamics in the case of free fermions and can efficiently handle non-local string operators arising from long-range interactions via transition matrix elements between non-orthogonal Slater determinants. 
We show that the fTDHF method can be implemented on a classical computer with a cost that scales polynomially with system size, 
and linearly with the time steps. 
We benchmark fTDHF against exact dynamics on three separate \spinhalf models, representing adiabatic preparation of states with long-range correlations, 
disorder-driven observation of many-body localization, and particle production in the Schwinger model. 
For each of these systems, fTDHF is shown to reproduce the qualitative dynamics generated by the exact evolutions, 
while maintaining a simple physical picture due to its mean-field nature. 

\end{abstract}


\maketitle


\section{Introduction} \label{sec: intro}

The Jordan--Wigner transformation (JWT) \cite{Jordan1928} is a celebrated technique for studying one-dimensional fermions and quantum spin systems. 
The JWT involves mapping \spinhalf operators into creation and annihilation operators of spinless fermions, combined with non-local diagonal operators known as JW strings. 
The JW strings ensure that the commutation relations for \spinhalf operators and the canonical anticommutation relations (CAR) for fermions are respected, 
a process commonly referred to as fermionization. \cite{Derzhko2008}
Fermionization is part of a broader physics principle, where finding dualities between different operator languages can help design simpler solutions to quantum many-body problems. \cite{Batista2004,Henderson2022duality}
Interestingly, most past applications of fermionization focus only on the scenarios where the JW strings are completely canceled out, if not ignored based on a physical justification. \cite{Grinberg2003,Gebhard2022localization} 
Nevertheless, this restriction still led to pioneering examples where certain \spinhalf systems with seemingly non-trivial interactions were mapped to Hamiltonians of 
non-interacting spinless fermions, which are exactly solvable, also known as free fermions.~\cite{Lieb1961, Katsura1962, yang1967, baxter1982, jimbo1989, sutherland2004, Derzhko2008}
In quantum computing, a class of classically simulable qubit-based unitaries, known as matchgates, is shown to be generated from free-fermion Hamiltonians. \cite{Valiant2001,Knill2001,Terhal2002} 

In this article, we focus on the quantum dynamics of a class of one-dimensional \spinhalf systems which are not exactly solvable due to the explicit presence of JW strings after fermionization. 
This includes linear spin chains with a wide variety of long-range interactions among the sites. 
Understanding Hamiltonians with long-range interactions is crucial across a wide range of fields, including condensed matter, \cite{Yoshioka:2022rej} 
nuclear and high-energy physics,~\cite{Bauer:2022hpo,Beck:2023xhh,Bauer:2023qgm,DiMeglio:2023nsa} 
and chemistry. \cite{RevModPhys.92.015003,Alexeev:2025how}
One-dimensional spin chains with non-trivial interactions are especially critical for emerging quantum technologies and engineering. 
For example, long-range XY models emerge from the resonant dipole-dipole interactions between Rydberg atoms, \cite{Browaeys2016experimental} 
and have been prepared on quantum simulators composed of trapped ions \cite{Maier2019environment,Feng2023continuous} and polar molecules. \cite{Yan2013observation}
A similar model has been shown to be the effective Hamiltonian for studying interactions between atoms and radiation fields. \cite{AsenjoGarcia2017}
A chain of XXZ spins with long-range interactions has also been experimentally simulated with trapped ions \cite{Hauke2010complete} and polar molecules. \cite{Hazzard2013magnetism}

The JW strings seem highly non-trivial from a computational perspective due to their non-local structure, which may have influenced their exclusion in earlier works. 
However, Scuseria and co-workers have recently shown in a series of articles \cite{Henderson2022duality, Henderson2024restoring, Henderson2024fermionic, Tabrizi2025scalable, Henderson2025strong, Tabrizi2026scf}
that these string operators can be computationally manageable if viewed as a special form of unitary Thouless rotation, \cite{Thouless1960} which is a widely used technique in quantum chemistry. \cite{HelgakerBook, BartlettBook}
Specifically, the action of the JW strings leads to unitary transformations of the underlying one-particle fermionic basis and transformations between non-orthogonal Slater determinants (SDs). \cite{Henderson2022duality}
These developments have been applied to study stationary states, especially the ground states, of spin and fermion-pair systems. 

In this work, we introduce a method where fermionization with explicit JW strings is applied to develop approximate quantum dynamics methods for spin systems beyond exactly solvable free-fermion models. 
Specifically, we develop a framework where a \spinhalf system is mapped to a Hamiltonian of spinless fermions, followed by a time-dependent Hartree--Fock (TDHF) approach to the unitary propagation which can handle non-local JW strings. 
For that, we assume that the state representing the dynamics of the fermionized system at each time step remains an SD, or in other words, a fermionic mean-field state.
We call this new method fermionized TDHF (fTDHF), which is formally equivalent to exact dynamics in the case of free fermions, and is potentially an excellent approximation for general scenarios, especially when the fermionized Hamiltonian has a dominant contribution from its free-fermionic part. 

The fTDHF method can be implemented on a classical computer with time that scales polynomially with system size, and linear with the time steps chosen. 
Since fTDHF applies to generic couplings for one-dimensional systems, it can be adapted to \spinhalf systems in a higher spatial dimension and with non-trivial geometric structures 
by simply enumerating them as a one-dimensional system with the corresponding set of long-range interactions. 
Efforts tailored to \spinhalf dynamics based on quantum computers are also evolving,
\cite{Wiebe2011simulating, Yao2021adaptive, kokcu2022, Peng2022time, bassman2022, Kim2023evidence, Gulania2024hybrid, Pocrnic2024qdrift, Gulania2025quantum}
where long-range interactions are often tackled via SWAP gates combined with quantum circuits for free fermions. \cite{kokcu2022, Gulania2024hybrid}

The rest of the article is organized as follows.
We review JWT and traditional TDHF in \Sec{\ref{sec: background}}. 
The implementation of fTDHF is discussed in \Sec{\ref{sec: ftdhf}}. 
We demonstrate the application of fTDHF on three separate \spinhalf Hamiltonians in \Sec{\ref{sec: results}}, 
focusing on adiabatic state preparation, many-body localization, and the Schwinger model. 
We finally conclude with discussions and future directions in \Sec{\ref{sec: final}}.


\section{Background} \label{sec: background}

We review the background concepts and establish the notations here. 
Following the discussion of the JWT and introducing the fermionized form of the spin Hamiltonian, we present the traditional TDHF method in a second quantization formulation. 

\subsection{Fermionization} 

Consider a \spinhalf system with $M$ sites, which the JWT maps into a system of $M$ spinless fermions, referred to as spin-orbital operators in quantum chemistry. 
Let us use $ \{ \FC{p}, \FA{p} \} $ to denote the fermionic creation and annihilation operators, respectively, which satisfy the CAR
\begin{equation}
\{ \FC{p}, \FA{q} \} 
= \delta_{pq}, \quad 
\{ \FC{p}, \FC{q} \} 
= 0.
\end{equation}
We also use $ \FN{p} = \FC{p} \FA{p} = \FN{p}^2 $ to denote the spin-orbital number operator, whose eigenvectors represent the fermionic occupations $\ket{0}$ and $\ket{1}$.
The commutation relation below will be useful later,
\begin{equation} \label{eq: num_op_comm}
[ \FN{p}, \FC{q} ] 
= \delta_{pq} \: \FC{q}.
\end{equation}
The JWT defines the mapping below \cite{Jordan1928}
\begin{subequations} \label{eq: jwt}
\begin{align}
\SP{p} 
&\mapsto \FC{p} \: \FS{p}, 
\\
\SM{p} 
&\mapsto \FS{p}^\dagger \: \FA{p},
\\
\SZ{p} 
&\mapsto \FN{p} - \frac{1}{2}, 
\end{align}
\end{subequations}
where $ \SP{p} = \big( X_p -  \mi \: Y_p \big) / 2 $, 
$ \SM{p} = \big( \SP{p} \big)^\dagger $, and
$ \SZ{p} = - Z_p / 2 $ are defined via the Pauli matrices for each site index $p$. 
The JW string operator $\FS{p}$, which we will simply call the string operator, is a non-local unitary operator
\begin{equation} \label{eq: jw_string}
\FS{p}
= \prod_{q = 1}^{p - 1} \: e^{ \mi \pi \FN{q} }
= \prod_{q = 1}^{p - 1} \: ( 1 - 2 \: \FN{q} ) 
= \prod_{q = 1}^{p - 1} \: \FNbar{q},
\end{equation}
where we have defined $ \FNbar{p} = 1 - 2 \: \FN{p} $. 
By virtue of the relation below
\begin{equation}
e^{ \mi \theta \: \FN{p} }
= 1 + \big( e^{\mi \theta} - 1 \big) \: \FN{p},
\end{equation}
the choice of either $ \theta = \pm \pi $ gives us the same operator, 
$ e^{ \mi \pi \: \FN{p} } = \FNbar{p} $, which means 
$ \FS{p}^\dagger = \FS{p} $.
It is clear from Eqs.~\eqref{eq: num_op_comm} and \eqref{eq: jw_string} that any $\FC{q}$ commutes with $\FS{p}$ if $q \geq p$.  
Even though we will focus only on the mapping of \Eq{\ref{eq: jwt}} in this work, the inverse mapping also exists and is widely used in the simulation of fermionic systems on qubit-based quantum computers. \cite{Chien2026simulating}
Note that the definitions here imply that the one-fermion state $\ket{0}$ is mapped to the spin-up state $\ket{\uparrow}$, and $\ket{1}$ to 
$\ket{\downarrow}$.

The tools we describe in this work can be applied to any \spinhalf Hamiltonian $H_S$ with the only restriction that it has global $\SZ{}$ symmetry
\begin{equation}
[ H_S, \SZ{} ] 
= \sum_{ p \: \in \: \text{all} } \: [ H_S, \SZ{p} ] 
= 0.
\end{equation}
This will make sure the mapped fermionic Hamiltonian $H_F$ conserves fermionic particle number, which allows us to apply Hartree--Fock (HF) techniques. 
It is possible to extend our methods to the most generic \spinhalf systems via fermionization by applying tools from the 
Hartree--Fock--Bogoliubov (HFB) theory\cite{Hu2014matrix, Kaicher2023gaussian, Chen2023robust} 
and its parity-breaking extensions,\cite{Colpa1979, Jafarizadeh2022parity, Henderson2024bogoliubov, Seifi2024generalization, Lyu2025displaced, Kaicher2026parity} which we will not explore in this paper. 
Given this restriction, a generic $H_S$ with up to two-body terms can be written as \cite{Henderson2022duality} 
\begin{subequations} \label{eq: generic_spin_ham}
\begin{align} 
H_S 
&= H_S^{(1)} + H_S^{(2)} + H_S^{(3)},
\\
H_S^{(1)} 
&= \sum_p \: J_p^{(1)} \: \SZ{p}
+ \sum_{p < q} \: J_{pq}^{(2)} \: \SZ{p} \SZ{q},
\\
H_S^{(2)} 
&= \sum_{p < q} \:J_{pq}^{(3)} \: \big( \SP{p} \SM{q} + \HermConj \big),
\\
H_S^{(3)}
&= \mi \: \sum_{p < q} \: J_{pq}^{(4)} \: \big( \SP{p} \SM{q} - \HermConj \big),
\end{align}
\end{subequations}
where $ \{ J_p^{(1)}, J_{pq}^{(2)}, J_{pq}^{(3)}, J_{pq}^{(4)} \} $ are real-valued Hamiltonian coefficients. 
The Hamiltonian in \Eq{\ref{eq: generic_spin_ham}} includes a wide range of spin systems, including the well-known XY and XXZ Heisenberg models. 

For conceptual clarity, we will first focus on a $H_S$ that is simpler than \Eq{\ref{eq: generic_spin_ham}} but still has a structure that highlights the novelty of our approach. 
Thus, we will choose the $H_S$ below from now on unless mentioned otherwise
\begin{equation} \label{eq: xy_all_ham}
H_S = \sum_{p < q} \: J_{pq} \: \Big( \SP{p} \SM{q} + \HermConj \Big), 
\end{equation}
which is the XY model with all-to-all and non-uniform interactions, i.e., each of the the $ J_{pq} = J_{qp} $ coefficient pairs can be arbitrary, and we assume all $J_{pp} = 0$.
We choose the Hamiltonian in \Eq{\ref{eq: xy_all_ham}} since the anti-symmetric terms in $H_S^{(3)}$ from \Eq{\ref{eq: generic_spin_ham}} can be similarly treated for the string operators, and 
the $\SZ{}$ terms in $H_S^{(1)}$ do not add any complication in this regard.
We refer the reader to \Appx{\ref{app: comm_matrix_expressions}} for discussions on handling the generic case defined by the Hamiltonian in \Eq{\ref{eq: generic_spin_ham}}.
Let us now map $H_S$ to the fermionic Hamiltonian,
\begin{align} \label{eq: xy_all_ham_fermi}
H_S \mapsto H_F
&= \sum_{p < q} J_{pq} \: \Big( 
\FC{p} \: \FS{p} \FS{q} \: \FA{q} + \HermConj \Big)
\nonumber 
\\
&= \sum_{p < q} J_{pq} \: 
\Big( \prod_{r = p + 1}^{q - 1} \FNbar{r} \Big) \: \big( \FC{p} \FA{q} + \HermConj \big),
\end{align}
where we have used the relation $ \FNbar{p}^2 = 1 $. 
It is clear that without the string operators, \Eq{\ref{eq: xy_all_ham_fermi}} represents a free-fermion system, which is precisely what happens for the nearest-neighbor-only interactions. 

\subsection{Time-Dependent Hartree--Fock} \label{sec: tdhf}

Let us consider a time-independent operator $\mathbb{O}$ and a many-fermion state $\ket{ \Psi(t) }$ evolving with time according to the Schr{\"o}dinger equation. 
Then we can write the time-evolution of the expectation value of $\mathbb{O}$ as
\begin{equation}
\frac{d}{dt} \braket{ \Psi(t) | \mathbb{O} | \Psi(t) }
= \mi \braket{ \Psi(t) | \: [ H_F, \mathbb{O} ] \: | \Psi (t)},
\end{equation}
where we have used the Schr{\"o}dinger equation 
\begin{equation}
\mi \: \frac{d}{dt} \ket{ \Psi(t) } 
= H_F \ket{ \Psi(t) },
\end{equation}
and $H_F$ is the many-fermion Hamiltonian of interest. 
In TDHF, we assume that the normalized state $\ket{ \Psi(t) }$ is a single SD at all times, which we denote as $\ket{ \psi(t) }$, 
and the one-particle reduced density matrix (1-RDM) elements are defined as
\begin{equation} \label{eq: 1-rdm}
\gamma_{pq}(t)
= \braket{ \psi(t) | \: \FC{p} \FA{q} \: | \psi(t) },
\end{equation}
where the $\{ \FC{p}, \FA{p} \}$ operators correspond to the one-particle basis states $\{ \ket{\varphi}_p \}$, which we call the canonical spin-orbitals.
The SD $\ket{ \psi(t) }$, on the other hand, is defined to be the Fermi vacuum constructed from $\{ \ket{ \chi_j} \} $, 
which is a set of orthonormal time-dependent one-particle basis represented as 
\begin{equation}
\ket{ \chi_j (t) } 
= \sum_{p \: \in \: \text{all}} C_{p, j} (t) \: \ket{ \varphi_p },
\end{equation}
where \textbf{C} is a $M \times M$ unitary matrix at all times.
In other words, we can write
\begin{equation}
\ket{\psi (t)}
= \Big( \prod_{j \: \in \: \text{occ}} \RFC{j} \Big) \ket{-},
\end{equation}
where ``occ'' denotes $N (\leq M)$ occupied spin-orbitals, and $ \ket{-} $ is the physical vacuum, also known as the true Fock vacuum.
We have defined the fermionic operators in the rotated basis as
\begin{equation}
\RFC{j} 
= \sum_{p \: \in \: \text{all}} 
C_{p, j} (t) \: \FC{p}, 
\end{equation}
where $ \{ \RFC{j}, \RFA{j} \} $ correspond to the $\{ \ket{ \chi_j } \} $ states, which also follow the CAR 
\begin{equation}
\{ \RFC{p}, \RFA{q} \} 
= \delta_{pq}, \quad 
\{ \RFC{p}, \RFC{q} \} 
= 0.
\end{equation}
Since we focus on unitary transformations of the spin-orbital bases, the inverse relation is straightforward,
\begin{equation}
\FC{p}
= \sum_{j \: \in \: \text{all}} 
C_{p, j}^* (t) \: \RFC{j}. 
\end{equation}
Note that any annihilation operator, rotated or otherwise, annihilates the physical vaccum 
\begin{equation}
\FA{p} \ket{-}
= \RFA{p} \ket{-} 
= 0.
\end{equation}
This means we can write \Eq{\ref{eq:  1-rdm}} as
\begin{align} \label{eq: 1-rdm_coeff}
\gamma_{pq} (t)
&= \sum_{j, k \: \in \: \text{all}} 
C_{p, j}^* (t) \: C_{q, k} (t) 
\braket{ \psi(t) | \: \RFC{j} \RFA{k} \: | \psi(t) }
\nonumber 
\\
&= \sum_{j, k \: \in \: \text{occ}} 
C_{p, j}^* (t) \: C_{q, k} (t) 
\braket{ \psi(t) | \: \big( \delta_{jk} 
+ \RFA{k} \RFC{j} \big) \: | \psi(t) }
\nonumber 
\\
&= \sum_{j \: \in \: \text{occ}} 
C_{p, j}^* (t) \: C_{q, j} (t),
\end{align}
where we used $\RFA{k} \ket{ \psi(t) } = 0$ for all unoccupied orbitals.
Thus, the $M \times M$ matrix $\bm{\gamma}$ is Hermitian and $\bm{\gamma}^2 = \bm{\gamma}$, i.e., idempotent, and can be constructed in $\ComCom{M^2 N}$ time. 
The expectation value of any $n$-body operator for SDs can be expressed in terms of the 1-RDM elements via Wick's theorem. \cite{Wick1950, BlaizotBook, BartlettBook}
Thus, the central object of TDHF is the equation of motion below \cite{RingBook}
\begin{equation} \label{eq: tdhf_eom}
\frac{d}{dt} \: \gamma_{pq} (t)
= \mi \braket{ \psi(t) | \: [ H_F, \FC{p} \FA{q} ] \: | \psi(t) } 
= \mi \: V_{pq},
\end{equation}
or $ \bm{\dot{\gamma}} = \mi \: \mathbf{V} $ in matrix form. 
As mentioned above, the \textbf{V} matrix can be expressed as a function of $\bm{\gamma}$.


\section{Fermionized TDHF} \label{sec: ftdhf}

We discuss the key parts of the fTDHF method in this section: the computations of the commutator matrix elements from the TDHF equation of motion with explicit string operators, followed by the time propagation. 

We again focus on the XY model with all-to-all interactions defined in \Eq{\ref{eq: xy_all_ham}} before discussing the generalizations. 
Our goal is to treat the string operators from $H_F$ in \Eq{\ref{eq: xy_all_ham_fermi}} exactly when computing the commutator matrix elements from \Eq{\ref{eq: tdhf_eom}},
\begin{widetext} 
\begin{align} \label{eq: xy_all_comm}
V_{pq}
&= \sum_{r < s} J_{rs} \: \Big[ 
\braket{ \psi | 
\Big( \prod_{j = r + 1}^{s - 1} \FNbar{j} \Big) 
\big( \FC{r} \FA{s} + \HermConj \big) \: 
\FC{p} \FA{q} \: | \psi } 
- \braket{ \psi | \:  
\FC{p} \FA{q} \: 
\big( \FC{r} \FA{s} + \HermConj \big)  
\Big( \prod_{j = r + 1}^{s - 1} \FNbar{j} \Big) 
| \psi } \Big]
\nonumber 
\\
&= \sum_{r < s} J_{rs} \: \Big[ 
\braket{ \psi_{r \rightarrow, \leftarrow s} | \: 
\big( \FC{r} \FA{s} + \HermConj \big) \: 
\FC{p} \FA{q} \: | \psi } 
- \braket{ \psi | \:  
\FC{p} \FA{q} \: 
\big( \FC{r} \FA{s} + \HermConj \big) \: 
| \psi_{r \rightarrow, \leftarrow s} } \Big], 
\end{align}
\end{widetext}
where $\ket{ \psi }$ is shorthand for $\ket{ \psi(t) }$ from now on. 
We will achieve this using an approach similar to \Reference{\citenum{Henderson2022duality}}.
We have defined the modified state as 
\begin{equation}
\ket{ \psi_{r \rightarrow, \leftarrow s} }
= \Big( \prod_{j = r + 1}^{s - 1} \FNbar{j} \Big) \: \ket{\psi},
\label{eq:psi_string}
\end{equation}
which is also an SD.  
Let us justify this statement by first expanding the single operator $\FN{p}$ as
\begin{equation}
\FN{p}
= \sum_{j, k \: \in \: \text{all}} 
C_{p, j}^* \: C_{p, k} \: \RFC{j} \RFA{k}
= \sum_{j, k \: \in \: \text{all}} 
G_{j, k}^{(p)} \: \RFC{j} \RFA{k},
\end{equation}
where the set of $M \times M$ matrices $ \{ \mathbf{G}^{(p)} \}$ are each Hermitian and idempotent.
The Thouless theorem states that the exponential of any one-body fermionic operator simply transforms one SD into another, \cite{Thouless1960} 
which means the unitary below
\begin{equation}
\FNbar{p}
= e^{ \mi \pi \: \FN{p} }
= \exp \Big( 
\mi \pi \sum_{j, k \: \in \: \text{all}} 
G_{j, k}^{(p)} \: \RFC{j} \RFA{k} \Big),
\end{equation}
rotates the underlying spin-orbital basis as
\begin{subequations}
\begin{align}
\ket{ \psi_{p} }
&= \FNbar{p} \ket{\psi}
\equiv \Big( \prod_{a \: \in \: \text{occ}} \:
\tilde{f}_{p, a}^\dagger \Big) \ket{-},
\\
\tilde{f}_{p, a}^\dagger
&= \sum_{j \: \in \: \text{all}} 
T_{j, a}^{(p)} \: \RFC{j}, 
\end{align}
\end{subequations}
where we have defined the $M \times M$ unitary matrix $\mathbf{T}^{(p)} 
= e^{i \pi \: \mathbf{G}^{(p)}} = 
\EYE - 2 \: \mathbf{G}^{(p)} $ due to its projector property.
This leads to the following relation
\begin{align}
\tilde{f}_{p, a}^\dagger
&= \sum_{j \: \in \: \text{all}} 
T_{j, a}^{(p)} \: \Big( 
\sum_{q \: \in \: \text{all}} 
C_{q, j} \: \FC{q} \Big)
\nonumber
\\
&= \sum_{j, q \: \in \: \text{all}} 
T_{j, a}^{(p)} \: C_{q, j} \: \FC{q} 
\nonumber
\\
&= \sum_{q \: \in \: \text{all}} 
\tilde{C}_{q, a}^{(p)} \: \FC{q}, 
\end{align}
where $\mathbf{C}$ gets modified by the single operator $\FNbar{p}$ as
\begin{equation}
\tilde{C}_{q, a}^{(p)}
= \sum_{j \: \in \: \text{all}} 
T_{j, a}^{(p)} \: C_{q, j}, 
\end{equation}
or in matrix form, 
$ \tilde{\mathbf{C}}^{(p)} = \mathbf{C} \: \mathbf{T}^{(p)} $.
This means that the coefficient matrix 
$ \mathbf{C} $ for $\ket{\psi}$ is updated to  
$ \tilde{\mathbf{C}}^{(p)} $ for 
$ \ket{ \psi_{p} } $ after applying the string operator $\FNbar{p}$ on it. 
At the next step, we can consider
$ \ket{ \psi_{p} } $ as the new reference SD and continue. 
This allows us to apply this procedure to 
$ \ket{ \psi_{r \rightarrow, \leftarrow s} } $ 
by iteratively applying each string operator with updated reference coefficient matrices, 
$ \mathbf{C} \Leftarrow \mathbf{\tilde{C}}^{(j)} $,
where $j$ runs from $r + 1$ to $s - 1$ following \Eq{\ref{eq:psi_string}}. 
We then denote $\tilde{\mathbf{C}}$ as the final updated coefficient matrix.

Thus, computation of \Eq{\ref{eq: xy_all_comm}} boils down to computing the transition density matrix elements of the form 
\begin{align}
&\braket{\psi | \: \FC{p} \FA{q} \: \FC{r} \FA{s} \: | \tilde{\psi} } 
= \braket{\psi | \: \FC{p} \: \big( 
\delta_{rq} - \FC{r} \: \FA{q} \big) \: \FA{s} \: | \tilde{\psi} } 
\nonumber
\\
&= \delta_{rq} \braket{\psi | \: \FC{p} \FA{s} \: | \tilde{\psi} } 
- \braket{\psi | \: \FC{p} \FC{r} \FA{q} \FA{s} \: | \tilde{\psi} }, 
\end{align}
where we have used the CAR, and 
$ \ket{\tilde{\psi}} = \ket{\psi_{r \rightarrow, \leftarrow s}} $ 
is non-orthogonal to the initial $\ket{\psi}$.
Different strategies have been developed to compute transition matrix elements between non-orthogonal SDs efficiently. \cite{Lowdin1955part1, Lowdin1955part2, RingBook, Hendekovic1981, JimenezHoyos2012canonical, Utsuno2013efficient, RodriguezLaguna2020efficient, Burton2021generalized}
Building the full \textbf{V} matrix with an efficient approach has a $\ComCom{M^4 N}$ cost, even with the generic Hamiltonian defined in \Eq{\ref{eq: generic_spin_ham}}, 
assuming the overlap matrix corresponding to $ \braket{\psi | \tilde{\psi} } $ is full-rank. 
In case of rank-deficient overlap matrices, the dominant cost increases to $\ComCom{M^4 N^2}$, as long as this only happens for a few cases of $\ket{\tilde{\psi}}$.
We refer the reader to \Appx{\ref{app: comm_matrix_expressions}} for more details on the expressions relevant to the \textbf{V} matrix, and 
to \Appx{\ref{app: tran_mat_elements}} for more details on computing the transition matrix elements between non-orthogonal SDs via computations involving
determinants and singular value decomposition (SVD).

Having established how to evaluate the matrix element $V_{pq}(t)$ in the SD framework, we now turn to the time integration 
of the resulting equations of motion defined in \Eq{\ref{eq: tdhf_eom}}. 
We discretize time with a uniform grid, $ t_n = n \: \Delta t$, and propagate the 1-RDM, $\bm{\gamma}(t)$, with the fourth-order Runge--Kutta (RK4) method. \cite{Butcher1996}
Denoting the function
\begin{equation}
F [ \bm{\gamma}(t) ] 
= \mi \: \mathbf{V}(t) \ ,
\end{equation}
the RK4 update from $t_n$ to $t_{n+1}=t_n+\Delta t$ is
\begin{align}
    \bm{k}_1 
    &= F[\bm{\gamma}_n],
    \nonumber \\
    \bm{k}_2 
    &= F[\bm{\gamma}_n + (\Delta t/2) \: \bm{k}_1],
    \nonumber \\
    \bm{k}_3 
    &= F[\bm{\gamma}_n + (\Delta t/2) \: \bm{k}_2], 
    \nonumber \\
    \bm{k}_4 
    &= F[\bm{\gamma}_n + ( \Delta t ) \: \bm{k}_3],  
    \nonumber \\
    \bm{\gamma}_{n+1} 
    &= \bm{\gamma}_n + \frac{\Delta t}{6} \: \big( 
    \bm{k}_1 + 2\: \bm{k}_2 + 2 \: \bm{k}_3 + \bm{k}_4 \big).
    \label{eq:RK4step}
\end{align}
Here, each evaluation of $F[\cdot]$ involves building the SD state $|\psi(t)\rangle$, or equivalently the coefficient matrix $\mathbf{C}$, 
and computing the matrix $\mathbf{V}$. 
Due to the multiple numerical steps of going back and forth between 1-RDMs, the coefficient matrix $\mathbf{C}$, and time integration, these matrices 
might accumulate numerical errors, i.e., loss of Hermiticity or misalignment of the orbitals. 
Details on how to numerically resolve these issues can be found in \Appx{\ref{app:projection}}.


\section{Results} \label{sec: results}

We have numerically verified that fTDHF reproduces the exact time evolution for free-fermionic cases.
We now apply fTDHF to three different systems where the long-range interactions are crucial to describe the phenomena observed, 
and compare its results against direct computations of the matrix exponentiation 
$ \ket{ \Psi(t) } = e^{- \mi t H_F} \ket{\Psi(0)} $ with a statevector representation on the Fock space.
We have verified that none of these three examples show any instance of rank-deficient orbital matrices, leading to the optimal scaling for fTDHF 
following the discussions in \Sec{\ref{sec: ftdhf}} and \Appx{\ref{app: tran_mat_elements}}.
Our numerical implementation used the Python libraries \texttt{SciPy} \cite{Virtanen2020scipy} and \texttt{OpenFermion}, \cite{Mcclean2020openfermion} 
and is openly available.~\cite{github}

\subsection{Adiabatic State Preparation with Long-Range Order}

The first example is the creation of long-range correlations in a one-dimensional system with long-range interactions,~\cite{Gong:2015nkd,PhysRevLett.119.023001} 
as it was experimentally realized in \Reference{\citenum{Feng2023continuous}}. 
Such states can be prepared by adiabatically evolving from a staggered-field to a long-range XY Hamiltonian,
\begin{align}
H (t) 
=&\  \frac{s(t)}{2} \: \sum_{ p < q }^{M} J_{pq} \: \big( \SP{p} \SM{q} + \HermConj \big) 
\nonumber
\\
&+ \big[ 1 - s(t) \big] \: \sum_{p = 1}^{M} h_p \: \SZ{p},
\end{align}
where the coefficients used are
$ J_{pq} = 0.28 \cdot e^{-0.19(q-p-1)}/(q-p)^{0.44}$ and $ h_p = 11.3 \cdot (-1)^p $, following \Reference{\citenum{Feng2023continuous}}.
The function $s(t)$ is the adiabatic ramp that goes from $0$ to $1$ for $t\in\{0,T=10\}$, and we take it to be an exponential, $s(t)=1-e^{-4t/T}$. 
If we start from the ground state of the staggered-field Hamiltonian, the final state belongs to the antiferromagnetic (XY) phase, 
displaying short-range correlations. 
However, if we start from the highest excited state of the staggered-field Hamiltonian, we obtain a ferromagnetic state 
that belongs to the continuous symmetry breaking (CSB) phase, with positive long-range spin correlations.

\begin{figure}[!t]
    \centering
    \includegraphics[width=\linewidth]{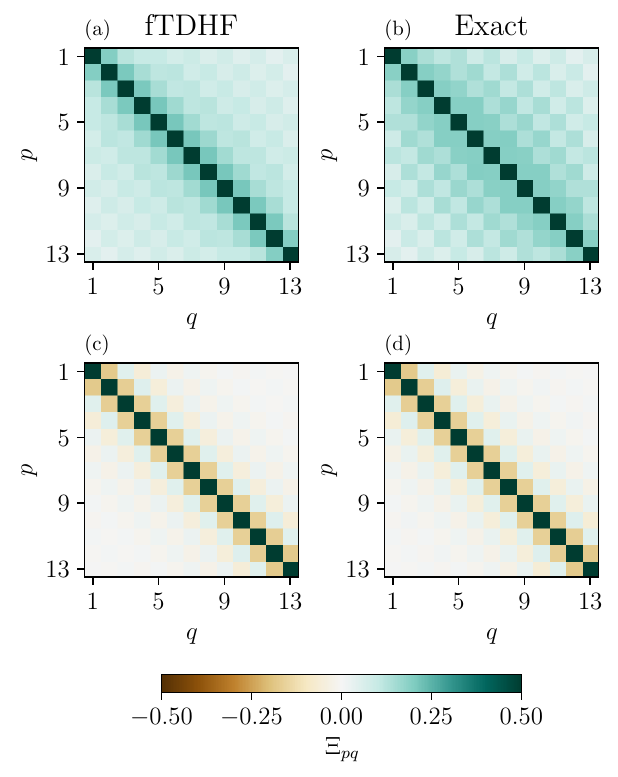}
    \caption{Spin-spin correlation matrix elements $\Xi_{pq} $ obtained using the fTDHF method (left) and with exact time evolution (right) for the state that belongs to the CSB phase, shown in panels (a) and (b), and for the state that belongs to the XY phase, shown in panels (c) and (d).}
    \label{fig:CpqMatrix}
\end{figure}

\begin{figure}[!t]
    \centering
    \includegraphics[width=\linewidth]{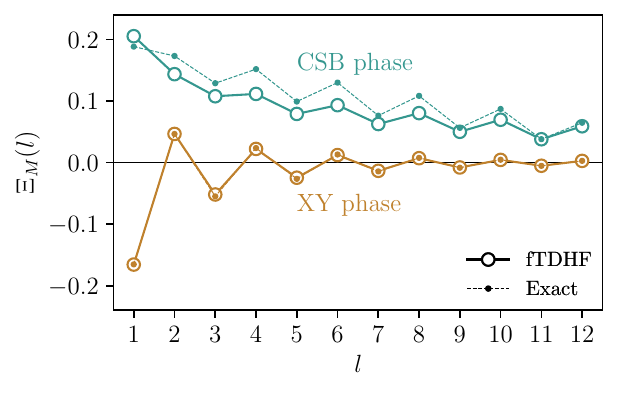}
    \caption{Spatially averaged spin-spin correlations $\Xi_M(l)$ as a function of distance $l$ obtained using the fTDHF method (open markers) and with exact time evolution (filled markers) for the states that belong to the CSB and XY phases.}
    \label{fig:CnAverage}
\end{figure}

We compute the spin-spin correlation matrix elements
$\Xi_{pq} = \braket{ \SP{p} \SM{q} + \SM{p} \SP{q} } $ after the time evolution for $M = 13$ spins, as shown in \Fig{\ref{fig:CpqMatrix}}. Since $\Xi_{pq}$ has a similar structure to Eq.~\eqref{eq: xy_all_comm}, the same techniques can be used here to compute this observable.
We also compare the spatially averaged correlations,
\begin{equation}
\Xi_M (l) 
= \sum_{p=1}^{M-l} \: \frac{1}{ M - l } \: \Xi_{p, p+l}, 
\end{equation} 
as shown in \Fig{\ref{fig:CnAverage}}.
We see that the fTDHF works better at preparing the state that belongs to the XY phase than to the CSB phase, but still captures the features of both phases.




\subsection{Many-Body Localization in the Presence of Disorder}

\begin{figure}[!t]
    \centering
    \includegraphics[width=\linewidth]{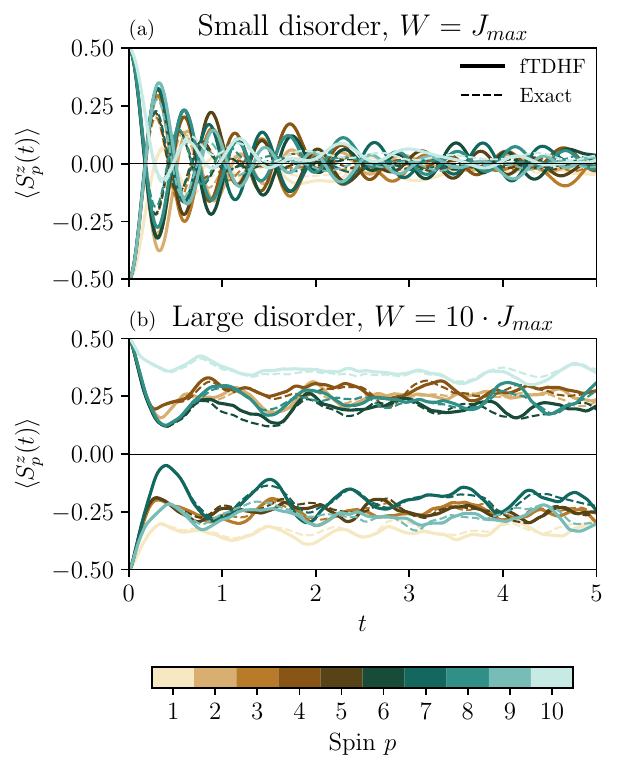}
    \caption{Expectation value of the $z$-spin alignment as a function of time, $\langle S^z_p(t)\rangle$, with (a) small disorder strength, $W=J_{max}$, and (b) large disorder strength, $W=10\cdot J_{max}$. The fTDHF values are shown with solid lines, while exact time evolution results are shown with dashed lines.}
    \label{fig:Sz_disorder}
\end{figure}

The next system we explore is the observation of the many-body localization (MBL) phase~\cite{PhysRev.109.1492,nandkishore2015} in a spin system with long-range interactions. 
Two key features are required: the model has to be non-integrable,~\cite{2007PhRvB..76e2203B} and disorder has to be present.~\cite{PhysRev.109.1492} 
Following \Reference{\citenum{Smith:2015dpe}}, we study the time evolution of the N\'eel state in the $z$-basis under the following Hamiltonian,
\begin{equation}
    H = \sum_{ p < q }^{M} J_{pq} \: \big( \SP{p} \SM{q} + \HermConj \big) 
    + B \sum_{p=1}^{M} \SZ{p} 
    + \sum_{p=1}^{M} D_p \: \SZ{p},
\end{equation}
where $ J_{pq} = J_{max} / |p-q|^{1.1} $, $ B = 4 \: J_{max} $ (with $J_{max}=\pi$), and $D_p$ is the disorder parameter, sampled 
from a uniform distribution, $D_p \in [-W,W]$. 
In the numerical experiment, we perform two runs with $M = 10$ spins, from small to large disorder, and look at the long-time value of the spin alignment $\braket{\SZ{p}(t)}$ (extracted from the diagonal elements of the 1-RDM) after averaging over 30 instances, as shown in \Fig{\ref{fig:Sz_disorder}}.


For the small disorder case, the spins relax to their thermal values, as expected by the eigenstate thermalization hypothesis (ETH).~\cite{PhysRevA.43.2046,PhysRevE.50.888} 
For the large disorder case, the state does not thermalize (ETH does not hold), with clear signatures of initial-condition memory effects on the magnetization of the spins.
The values obtained using fTDHF follow more closely the exact results when there is a large disorder in the system. 
This is justified from a perturbation perspective since the free-fermionic disorder term becomes the zeroth-order Hamiltonian in this case. \cite{Burin2015localization}
For the small disorder case, while differences appear earlier in time, both sets approach the same value, $ \braket{ \SZ{p} (t) } = 0 $.

\subsection{Electron-Positron Pair Creation in the Schwinger Model}

For the final example, we move from the condensed matter systems to the Schwinger model,~\cite{PhysRev.128.2425} which has become the benchmark system 
in quantum simulations for nuclear and high-energy physics.~\cite{Martinez:2016yna,Klco:2018kyo,deJong:2021wsd,Farrell:2023fgd,Farrell:2024fit,Davoudi:2024wyv} 
This model describes quantum electrodynamics in 1+1 dimensions, but shares many features with quantum chromodynamics in 3+1 dimensions, such as confinement and a non-trivial vacuum. 
The Hamiltonian on a staggered lattice of size $L$ (and $ M = 2 \: L$ spins) with open-boundary conditions can be written as~\cite{PhysRevD.11.395,Banuls:2013jaa}
\begin{align}
    H =&\ x \sum_{p = 1}^{M-1} \big( \SP{p} \SM{p+1} + \HermConj \big) 
      + \mu \sum_{p = 1}^{M} \Big[ \frac{1}{2} - (-1)^{p+1} \: \SZ{p} \Big] 
      \nonumber \\
      &+ \sum_{p = 1}^{M-1} \Big( \sum_{q=1}^{p} \big[ (-1)^{q+1} \: \frac{1}{2} - \SZ{q} \big] \Big)^2,
\label{eq:ScwhingerHam}
\end{align}
with $x=1/g^2a^2$, $ \mu = 2 \: \sqrt{x} \: (m/g) $, $m$ being the electron mass, $g$ the coupling constant and $a$ the lattice spacing. 
We note that 
non-Abelian gauge theories in 1+1 dimensions have a similar Hamiltonian to Eq.~\eqref{eq:ScwhingerHam}, hence fTDHF can also be applied.~\cite{Farrell:2022wyt}


%
\begin{figure}[!t]
    \centering
    \includegraphics[width=\linewidth]{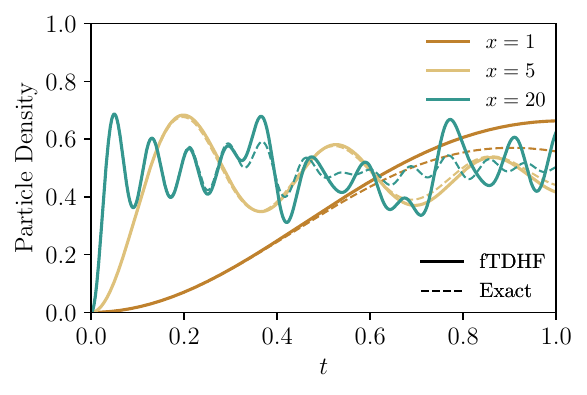}
    \caption{Evolution of the particle density as a function of time for different values of $x$, as defined below \Eq{\ref{eq:ScwhingerHam}}, starting from the bare vacuum. 
    The fTDHF values are shown with solid lines, while exact time evolution results are shown with dashed lines.}
    \label{fig:scwhinger}
\end{figure}
For demonstration purposes, we study a lattice with $L=6$ sites ($M=12$ spins), fixing the ratio $m/g=0.125$ and tuning the parameter $x$ from $1$ to $20$ to approach the continuum limit 
(ideally, $L$ would also be increased to keep the spatial extent constant as we go to finer lattices). 
In particular, we study the dynamics of creation and annihilation of pairs of electron-positron from the bare vacuum (Néel state). 
We compute the particle density, as defined below
\begin{equation}
    \nu (t) 
    = \frac{1}{M} \: \sum_{p=1}^{M} \left[ \frac{1}{2}
    - (-1)^{p+1} \braket{ \SZ{p} } \right],
\end{equation}
and compare them for fTDHF and exact dynamics in \Fig{\ref{fig:scwhinger}}. 
As expected, fTDHF can reproduce the early-time dynamics accurately, where correlations beyond single SD have not yet built up.


\section{Discussions} \label{sec: final}

We have introduced a real-time quantum dynamics method called fTDHF for \spinhalf systems, inspired by techniques in areas related to 
many-fermion systems, such as electronic structure theory. 
Specifically, we have shown how a combination of JWT and Thouless rotations allows us to compute time-dependent expectation 
values efficiently in a mean-field setting. 
The fTDHF method can be implemented on a classical computer, and its computational cost scales polynomially with the system size.
Across the examples shown above, we find that fTDHF can capture the dominant dynamical signatures. Deviations from exact results appear in regimes where entanglement growth and higher-order correlators become significant, which is consistent with the single-SD ansatz underlying fTDHF.

The reasonable agreement of fTDHF with exact dynamics, as shown in \Sec{\ref{sec: results}}, can be justified from the perspective of the \spinhalf operators. 
Since $\ket{\psi (t)}$ at any time $t$ is assumed to be an SD represented in a different spin-orbital basis, 
it can be written as
\begin{equation}
\ket{\psi (t)}
= e^{K_1} \ket{\psi (0)}, \quad 
K_1 
= \sum_{p < q} \: \kappa_{pq} \: \big( \FC{p} \FA{q} - \HermConj \big),
\end{equation}
where $K_1$ can be mapped to a spin operator via the inverse JWT as 
\begin{equation}
K_1 
\mapsto \sum_{p < q} \kappa_{pq} \: \Big( \prod_{r = p + 1}^{q - 1} Z_r \Big) \: 
\big( \SP{p} \SM{q} - \HermConj \big).
\end{equation}
Thus, if we assume $\ket{\psi (0)}$ is a computational basis product state, 
$\ket{\psi (t)}$ is a non-trivial superposition of spin basis states, even though $e^{K_1}$ is not universal. 
Thus, fTDHF is expected to perform well for systems beyond free fermions for which $e^{K_1}$ is expressive enough.
 
This work opens the door for exploring several future directions, which we mention below. 
As discussed in \Sec{\ref{sec: background}}, fTDHF assumes the Hamiltonian of interest has global $\SZ{}$ symmetry, which can be extended to 
arbitrary \spinhalf Hamiltonians by generalizing the SDs to HFB states. 
This work can be extended to study higher-dimensional \spinhalf systems by flattening them into one dimension, at the price of 
increasing non-local interactions, which motivate future developments to lower the computational costs of the expectation values. 
We have demonstrated the time evolution in \Sec{\ref{sec: results}} with generic integrators, which can have issues maintaining the 
fermionic structure of the 1-RDM.
\Appx{\ref{app:projection}} discusses fixing the 1-RDMs numerically, but it may be more convenient to apply integrators that 
are naturally structure-preserving, \cite{Kosloff1988time, Castro2004propagators, Gomez2018propagators, Li2020real}
after building an analogue of the Fock matrix from standard TDHF. 
fTDHF involves no adjustable parameters and operates at a fixed level of approximation, with a clear physical justification based on the time-evolved SDs after the fermionization step.  
This is complementary to the state-of-the-art quantum dynamics methods based on classical computing, where the accuracy can be systematically controlled to approach 
the exact dynamics with increasing computational cost. \cite{Cazalilla2002, White2004time, Vidal2004quantum, Kuprov2007polynomially, Haegeman2011time, Carleo:2016svm, Ma2018time, Patrick2019pauli, Begusic2024fast, Shao2024noisy, Thomson2024unravelling, Begusic2025sparse, Rudolph2025pauli} 
Thus, fTDHF can provide a physically inspired starting point for more involved quantum dynamical calculations, and we leave exploration on this front to future work. 


\begin{acknowledgments}

We acknowledge support from Pacific Northwest National Laboratory’s Quantum Algorithms and Architecture for Domain Science (QuAADS) Laboratory Directed Research and Development (LDRD) Initiative.
The authors also acknowledge support from the U.S. Department of Energy, Office of Science, National Quantum Information Science Research Centers, 
Quantum Science Center (R.~D., M.~I.\ and K.~K.\ under FWP 76213) and Q-NEXT (N.~G. under FWP 76155), at Pacific Northwest National Laboratory.

\end{acknowledgments}


\section*{Code and Data Availability}

The source code to reproduce the examples shown in the paper is available in a GitHub repository.~\cite{github}


\appendix


\section{Generic Commutator Matrix Element Expressions} \label{app: comm_matrix_expressions}

We discuss expressions for the \textbf{V} matrix from \Sec{\ref{sec: ftdhf}} here for the general case. 
The fermionic algebra expressions have been simplified with help from the \texttt{QuantumAlgebra.jl} library. \cite{QuantumAlgebraJulia}
The mapped fermionic Hamiltonian $H_F$ from the general $H_S$ defined in \Eq{\ref{eq: generic_spin_ham}} can be written as 
\begin{subequations} \label{eq: generic_fermi_ham}
\begin{align} 
H_F
&= H_F^{(0)} + H_F^{(1)} + H_F^{(2)} + H_F^{(3)},
\\
H_F^{(0)}
&= - \frac{1}{2} \: \sum_p \: J_p^{(1)} 
+ \frac{1}{4} \: \sum_{p < q} \: J_{pq}^{(2)},
\\
H_F^{(1)} 
&= \sum_p \: J_p^{(1)} \: \FN{p}
- \frac{1}{2} \: \sum_{p < q} \: J_{pq}^{(2)} \: \big( \FN{p} + \FN{q} \big)
\nonumber
\\
&+ \sum_{p < q} \: J_{pq}^{(2)} \: \FN{p} \FN{q},
\\
H_F^{(2)} 
&= \sum_{p < q} \: J_{pq}^{(3)} \: \Big( \prod_{r = p + 1}^{q - 1} \FNbar{r} \Big) \: 
\big( \FC{p} \FA{q} + \HermConj \big),
\\
H_F^{(3)}
&= \mi \: \sum_{p < q} \: J_{pq}^{(4)} \: \Big( \prod_{r = p + 1}^{q - 1} \FNbar{r} \Big) \: 
\big( \FC{p} \FA{q} - \HermConj \big).
\end{align}
\end{subequations}
It is clear that $ [ H_F^{(0)}, \FC{p} \FA{q} ] = 0 $, so let us now focus on the diagonal terms below
\begin{subequations}
\begin{align} 
[ \FN{r}, \FC{p} \FA{q} ] 
&= ( \delta_{pr} - \delta_{qr} ) \: \FC{p} \FA{q},
\\
[ \FN{r} \FN{s}, \FC{p} \FA{q} ] 
&= ( \delta_{pr} \delta_{ps} - \delta_{qr} \delta_{qs} ) \: \FC{p} \FA{q}
+ ( \delta_{qs} - \delta_{ps} ) \: \FC{p} \FC{r} \FA{q} \FA{r}
\nonumber
\\
&+ ( \delta_{qr} - \delta_{pr} ) \: \FC{p} \FC{s} \FA{q} \FA{s},
\end{align}
\end{subequations}
which leads to the following commutator 
\begin{align} 
[ H_F^{(1)}, \FC{p} \FA{q} ] 
&= \Big[ J_p^{(1)} - J_q^{(1)} 
+ \frac{1}{2} \: \sum_r \: \Big( J_{qr}^{(2)} - J_{pr}^{(2)} 
\Big) \Big] \: \FC{p} \FA{q} 
\nonumber
\\
&+ \sum_r \: \Big( J_{qr}^{(2)} - J_{pr}^{(2)} \Big) \: 
\FC{p} \FC{r} \: \FA{q} \FA{r},
\end{align}
where we have assumed $ J_{pq}^{(2)} = J_{qp}^{(2)} $ and all $ J_{pp}^{(2)} = 0 $.
We can now write 
\begin{align} 
\mathcal{T}_{pq}^{(1)}
=&\ \braket{ \psi| \: [ H_F^{(1)}, \FC{p} \FA{q} ] \: | \psi}
\nonumber
\\
=&\ \Big[ J_p^{(1)} - J_q^{(1)} 
+ \frac{1}{2} \: \sum_r \: \Big( J_{qr}^{(2)} - J_{pr}^{(2)} 
\Big) \Big] \: \gamma_{pq}
\nonumber
\\
&+ \sum_r \: \Big( J_{qr}^{(2)} - J_{pr}^{(2)} \Big) \: \Big( 
\gamma_{pr} \: \gamma_{rq}
- \gamma_{pq} \: \gamma_{rr} \Big).
\end{align}
Building the full $ \bm{ \mathcal{T} }_{M \times M}^{(1)}  $ matrix has 
$ \ComCom{M^3} $ cost after building the $\bm{\gamma}$ matrix in $\ComCom{M^2 N}$ time following \Sec{\ref{sec: tdhf}}.

Let us now simplify the following terms
\begin{subequations}
\begin{align}
&\big( \FC{r} \FA{s} + \FC{s} \FA{r} \big) \: \FC{p} \FA{q}
\nonumber
\\
&= \delta_{ps} \: \FC{r} \FA{q}
+ \delta_{pr} \: \FC{s} \FA{q}
- \FC{p} \FC{r} \: \FA{q} \FA{s}
- \FC{p} \FC{s} \: \FA{q} \FA{r},
\\
&\FC{p} \FA{q} \: \big( \FC{r} \FA{s} + \FC{s} \FA{r} \big)
\nonumber
\\
&= \delta_{qs} \: \FC{p} \FA{r}
+ \delta_{qr} \: \FC{p} \FA{s}
- \FC{p} \FC{r} \: \FA{q} \FA{s}
- \FC{p} \FC{s} \: \FA{q} \FA{r},
\end{align}
\end{subequations}
and then define the following tensor elements
\begin{subequations}
\begin{align}
R_{pq}^{(jk)}
&= \braket{ \psi | \: \FC{p} \FA{q} \: | \psi_{j \rightarrow, \leftarrow k} },
\\
\tilde{R}_{pq}^{(jk)}
&= \braket{ \psi_{j \rightarrow, \leftarrow k} | \: \FC{p} \FA{q} \: | \psi },
\\
Z_{pq rs}^{(jk)}
&= \braket{ \psi | \: \FC{p} \FC{q} \: \FA{s} \FA{r} \: | \psi_{j \rightarrow, \leftarrow k} },
\\
\tilde{Z}_{pq rs}^{(jk)}
&= \braket{ \psi_{j \rightarrow, \leftarrow k} | \: \FC{p} \FC{q} \: \FA{s} \FA{r} \: | \psi }.
\end{align}
\end{subequations}
Constructing $ \{ \bm{Z}^{(jk)}, \bm{\tilde{Z}}^{(jk)} \} $ dominate the cost among the intermediate tensors above, 
and has $\ComCom{1}$ cost for each $(j, k)$ pair after constructing $\bm{\rho}$ in $\ComCom{M^2 N}$ time following \Appx{\ref{app: tran_mat_elements}}.
Thus, the full construction of $ \bm{Z} $ and $ \bm{\tilde{Z}} $ has $\ComCom{M^4 N}$ cost. 
However, in the case where $ \braket{ \psi | \psi_{j \rightarrow, \leftarrow k} } $ corresponds to a numerically rank-deficient overlap matrix, 
constructing $ \{ \bm{Z}^{(jk)}, \bm{\tilde{Z}}^{(jk)} \} $ increases to $\ComCom{M^4 N^2}$, as discussed in \Appx{\ref{app: tran_mat_elements}}.
In the worst case, this can lead to an $\ComCom{M^6 N^2}$ cost for the full construction of $ \bm{Z} $ and $ \bm{\tilde{Z}} $, but in practice, 
the rank-deficient scenario may be encountered only for an $\ComCom{1}$ number of $(j, k)$ pairs, if at all.
Following \Sec{\ref{sec: ftdhf}}, we can write 
\begin{align} 
\mathcal{T}_{pq}^{(2)}
&= \braket{ \psi| \: [ H_F^{(2)}, \FC{p} \FA{q} ] \: | \psi}
\nonumber
\\
&= \frac{1}{2} \: \sum_r \: \Big[ J_{pr}^{(3)} \: \Big( 
\tilde{R}_{rq}^{(pr)} + \tilde{R}_{rq}^{(rp)} \Big)
\nonumber
\\
&+ J_{qr}^{(3)} \: \Big( R_{pr}^{(qr)} + R_{pr}^{(rq)} \Big) \Big] 
\nonumber
- \sum_{r < s} \: J_{rs}^{(3)} \: \Big[ 
\tilde{Z}_{pr sq}^{(rs)} 
\nonumber
\\
&- \tilde{Z}_{ps rq}^{(rs)}
+ Z_{pr sq}^{(rs)} - Z_{ps rq}^{(rs)} \Big].
\end{align}
Assuming all intermediate tensors defined above have been constructed, the full 
$ \bm{ \mathcal{T} }_{M \times M}^{(2)} $ matrix can be constructed in $ \ComCom{M^4} $ time. 
A similar analysis can be shown for the $ \braket{ \psi| \: [ H_F^{(3)}, \FC{p} \FA{q} ] \: | \psi} $ part. 


\section{Transition Matrix Elements Between Non-Orthogonal Slater Determinants} \label{app: tran_mat_elements}

In this appendix, we review the concepts and practical implementations for computing the transition matrix elements between two non-orthogonal SDs.
We have mostly followed the material discussed in \Reference{\citenum{Utsuno2013efficient}} and \Reference{\citenum{RodriguezLaguna2020efficient}}, unless mentioned otherwise. 

\subsection{Background}

Let us consider two SDs of $N$ spinless fermions
\begin{subequations}
\begin{align}
\ket{A_N} 
&= \FClt{1} \cdots \FClt{N} \ket{-},
\\
\ket{B_N} 
&= \FCrt{1} \cdots \FCrt{N} \ket{-},
\end{align}
\end{subequations}  
where  $\ket{-}$ is the physical vacuum, and $\braket{A_N | B_N} \neq 0$.
The $ \{ \FClt{j} \} $ and $ \{ \FCrt{j} \} $ are two sets of arbitrary creation operators with indices representing the one-particle states they create, with their corresponding Hermitian conjugates, the annihilation operators $ \{ \FAlt{j} \} $ and $ \{ \FArt{j} \} $, annihilate the physical vacuum
\begin{equation}
\FAlt{j} \ket{-}
= \FArt{j} \ket{-} = 0, 
\end{equation}
Each of the two sets of operators satisfies the CAR, and the anticommutation relations between the two sets are defined as 
\begin{equation} \label{eq: nonortho_anticomm}
\{ \FAlt{j}, \FCrt{k} \} 
= \FOv{jk}, \quad 
\{ \FClt{j}, \FCrt{k} \}
= 0,
\end{equation}
where we define the elements of the overlap matrix as
\begin{equation} \label{eq: overlap_mat_elems}
\FOv{jk} 
= \braket{ - | \: \FAlt{j} \: \FCrt{k} \: | - }, 
\end{equation}
and \textbf{S} is an $N \times N$ matrix. 

Let us now represent each of the two sets in terms of a common basis of $M \: (\geq N)$ dimensions 
\begin{subequations}
\begin{align}
\FClt{j}
&= \sum_{p = 1}^M \: A_{pj} \: \FC{p}, \quad
\forall \: j = 1, \dots, N,
\\
\FCrt{j}
&= \sum_{p = 1}^M \: B_{pj} \: \FC{p}, \quad
\forall \: j = 1, \dots, N,
\end{align}
\end{subequations}
where $\{ \FC{p} \}$ are the creation operators in the common basis with the same elementary characteristics as the previous ones, and $\mathbf{A}$ and $\mathbf{B}$ are $M \times N$ expansion matrices. 
Then we can write 
\begin{equation}
\FOv{jk}
= \sum_{pq} \: \braket{ - | \: \FA{p} \: \FC{q} \: | - } \: A_{pj}^* \: B_{qk}
= \sum_{p} \: A_{pj}^* \: B_{pk},
\end{equation}
where we have used 
$ \braket{ - | \: \FA{p} \: \FC{q} \: | - } = \delta_{pq} $.
In matrix form, 
$\mathbf{S}_{N \times N} = ( \mathbf{A}^\dagger )_{N \times M} \: \mathbf{B}_{M \times N} $, 
with $\ComCom{M N^2}$ cost is associated with building the full \textbf{S} matrix.
Let us also introduce the operators for the unoccupied states in the common basis as
\begin{subequations}
\begin{align}
\FClt{a}
&= \sum_{p = 1}^M \: \tilde{A}_{pa} \: \FC{p}, \quad
\forall \: a = N + 1, \dots, M,
\\
\FCrt{a}
&= \sum_{p = 1}^M \: \tilde{B}_{pa} \: \FC{p}, \quad
\forall \: a = N + 1, \dots, M,
\end{align}
\end{subequations}
where $L = M - N$ is the number of unoccupied spin-orbitals and $\mathbf{\tilde{A}}$ and $\mathbf{\tilde{B}}$ are the corresponding $M \times L$ expansion matrices.
We will use the indices $j, k$ and $a, b$ for the occupied and unoccupied states in this section, respectively, and the indices $p, q$ for the common basis states.
Then we can write
\begin{subequations}
\begin{align}
\FC{p}
&= \sum_{j = 1}^N \: A_{pj}^* \: \FClt{j}
+ \sum_{a = N + 1}^M \: \tilde{A}_{pa}^* \: \FClt{a},
\\
\FC{p}
&= \sum_{j = 1}^N \: B_{pj}^* \: \FCrt{j}
+ \sum_{a = N + 1}^M \: \tilde{B}_{pa}^* \: \FCrt{a}.
\end{align}
\end{subequations} 
Applying the CAR, we get the following relation 
\begin{equation}
\EYE_{M \times M} 
= \mathbf{A} \: \mathbf{A}^\dagger 
+ \mathbf{\tilde{A}} \: \mathbf{\tilde{A}}^\dagger 
= \mathbf{B} \: \mathbf{B}^\dagger 
+ \mathbf{\tilde{B}} \: \mathbf{\tilde{B}}^\dagger. 
\end{equation}
It can also be shown that 
\begin{subequations}
\begin{align}
\FA{p} \ket{B_N} 
&= \Big( \sum_{j = 1}^N \: B_{pj} \: \FArt{j} \Big) \ket{B_N},
\\
\bra{A_N} \FC{p}
&= \bra{A_N} \Big( \sum_{j = 1}^N \: A_{pj}^* \: \FClt{j} \Big),
\end{align}
\end{subequations} 
where we have used 
$ \FArt{a} \ket{B_N} = \bra{A_N} \FClt{a} = 0 $, 
which means we do not explicitly need $\mathbf{\tilde{A}}$ and $\mathbf{\tilde{B}}$ for the transition overlap expressions below.  

\subsection{Transition Matrix Elements}

The overlap can be written using \Eq{\ref{eq: nonortho_anticomm}} as
\begin{align} \label{eq: nonortho_overlap}
\braket{A_N | B_N }
=&\ \braket{ - | \: \FAlt{N} \cdots \big( 
\FAlt{1} \: \FCrt{1} \big) \cdots \FCrt{N} \: | - } 
\nonumber 
\\
=&\ - \braket{ - | \: \FAlt{N} \cdots \big( 
\FAlt{2} \: \FCrt{1} \big) \: \FAlt{1} \: \FCrt{2} \cdots \FCrt{N} \: | - } 
\nonumber
\\
&+ \FOv{11} \braket{ - | \: \FAlt{N} \cdots \big( 
\FAlt{2} \: \FCrt{2} \big) \cdots \FCrt{N} \: | - },
\end{align}
where the recursive relation above continues for $N$ steps until $ \bra{-} \FCrt{1} = 0 $. 
The $N = 1$ case is already defined as $S_{11}$.  
The $N = 2$ case is shown below 
\begin{align}
\braket{A_2 | B_2 } 
&= - \braket{ - | \: \big( \FAlt{2} \: \FCrt{1} \big) \: 
\FAlt{1} \: \FCrt{2} \: | - } 
+ \FOv{11} \: \FOv{22}
\nonumber
\\
&= \FOv{11} \: \FOv{22} - \FOv{12} \: \FOv{21}
\nonumber
\\
&= \det ( \mathbf{S}_{1:2, 1:2} ), 
\end{align} 
where $\det (\cdot) $ denotes the determinant of a square matrix, and we use the notation $j:k$ to represent matrix blocks. 
Thus, we can recursively expand \Eq{\ref{eq: nonortho_overlap}} as 
\begin{align} \label{eq: nonortho_overlap_expanded}
\braket{A_N | B_N }
=&\ \FOv{11} \braket{A_{N-1} | B_{N-1} }_{(1, 1)} 
\nonumber
\\
&- \FOv{21} \braket{A_{N-1} | B_{N-1} }_{(2, 1)} 
+ \cdots
\nonumber
\\
&+ (-1)^{N-1} \FOv{N1} \braket{A_{N-1} | B_{N-1} }_{(N, 1)},
\end{align} 
where the subscript $(p, q)$ indicates that the pair 
$ \FAlt{p} \FCrt{q} $ has been removed from 
$ \braket{A_N | B_N } $.
Based on \Eq{\ref{eq: nonortho_overlap_expanded}} and the $N = 2$ case, the overlap becomes
\begin{equation} \label{eq: nonortho_overlap_final}
\braket{A_N | B_N }  
= \det ( \mathbf{S} ),
\end{equation}
where the Laplace expansion of a determinant via its minors has been applied.
Since \textbf{S} has $N \times N$ dimensions, computation of \Eq{\ref{eq: nonortho_overlap_final}} has $\ComCom{N^3}$ cost. 

Let us now focus on the elementary contraction 
\begin{align} \label{eq: rho}
\rho_{pq}
&= \frac{ \braket{A_N | \: \FC{p} \: \FA{q} \: | B_N } }{ \braket{A_N | B_N } }
\nonumber
\\
&= \sum_{jk} \: A_{pj}^* \: B_{qk} \: 
\frac{ \braket{A_N | \: \FClt{j} \: \FArt{k} \: | B_N } }{ \det ( \mathbf{S} ) },
\end{align} 
where the dependence of $\bm{\rho}$ on the states $\ket{A_N}, \ket{B_N}$ is dropped for brevity.
Consider the action of $\FArt{k}$ on the $\ket{B_N}$ state 
\begin{equation}
\FArt{k} \ket{B_N}
= \FArt{k} \: \FCrt{1} \cdots \FCrt{N} \ket{-} 
= ( - 1)^{k + 1} \ket{ B_{N - 1}^{(k)} },
\end{equation}
where the superscript $(k)$ indicates that the $\ket{B_{N - 1}}$ state does not contain the $\FCrt{k}$ operator. 
Similarly, we can write 
\begin{equation}
\bra{A_N} \FClt{j}
= \bra{ A_{N - 1}^{(j)} } ( - 1)^{j + 1},
\end{equation}
which leads to the following relation
\begin{align}
\braket{A_N | \: \FClt{j} \: \FArt{k} \: | B_N }
&= ( - 1)^{j + k} \braket{A_{N - 1} | B_{N - 1} }_{(j, k)}
\nonumber
\\
&= ( - 1)^{j + k} \det \big[ \mathbf{S}_{(j, k)} \big], 
\end{align}
where $ \mathbf{S}_{(j, k)} $ is the $(N - 1) \times (N - 1)$ submatrix of $\mathbf{S}$ with the $j$-th row and $k$-th column removed.
Cramer's rule for matrices states that
\begin{equation} \label{eq: cramer_rule}
\big( \mathbf{S}^{-1} \big)_{kj}
= \frac{ ( - 1)^{j + k} \det \big[ \mathbf{S}_{(j, k)} \big] }{ \det ( \mathbf{S} ) },
\end{equation} 
which allows us to represent the $ \bm{\rho}_{M \times M} $ matrix from \Eq{\ref{eq: rho}} as
\begin{subequations} \label{eq: rho_final}
\begin{align}
\rho_{pq}
&= \sum_{jk} \: 
A_{pj}^* \: B_{qk} \: \big( \mathbf{S}^{-1} \big)_{kj},
\\
\bm{\rho}
&= \mathbf{A}^* \: \big( \mathbf{S}^{-1} \big)^T \: 
\mathbf{B}^T.
\end{align}
\end{subequations}
The cost of computing the full $\bm{\rho}$ matrix is $\ComCom{M^2 N}$.

Following \Reference{\citenum{Hendekovic1981}}, any transition matrix element involving number-conserving terms can be computed as a function of $\bm{\rho}$ via 
a generalized  Wick's theorem
\begin{align} \label{eq: gen_wick}
&\braket{A_N | \: \FC{p_1} \cdots \FC{p_d} \: \FA{q_d} \cdots \FA{q_1} \: | B_N }
\nonumber
\\
&= \det ( \mathbf{S} ) \: \det ( \bm{\Gamma}_d ),
\end{align}
where the elements of the $d \times d$ matrix $\bm{\Gamma}_d$ are
\begin{equation}
\big[ \bm{\Gamma}_m \big]_{\mu \nu}
= \rho_{p_\mu q_\nu}.
\end{equation}
For example, the two-body transition matrix elements can be written as 
\begin{equation} \label{eq: 2_trdm}
\braket{A_N | \: \FC{p} \: \FC{q} \: \FA{s} \: \FA{r} \: | B_N }
= \det ( \mathbf{S} ) \: ( \rho_{pr} \: \rho_{qs} - \rho_{ps} \: \rho_{qr} ).
\end{equation}
\Eq{\ref{eq: gen_wick}} reduces to the standard Wick's theorem for the case below 
\begin{equation} \label{eq: wick}
\braket{ \psi | \: \FC{p_1} \cdots \FC{p_d} \: \FA{q_d} \cdots \FA{q_1} \: | \psi }
= \det ( \bm{\Gamma}_d^{(0)} ),
\end{equation} 
where the elements of the $d \times d$ matrix $\bm{\Gamma}_d^{(0)}$ are
\begin{equation}
\big[ \bm{\Gamma}_d^{(0)} \big]_{\mu \nu}
= \braket{ \psi | \: \FC{p_\mu} \FA{q_\nu} \: | \psi }
= \gamma_{p_\mu q_\nu},
\end{equation}
and $\ket{\psi}$ is an arbitrary normalized SD.

\subsection{Tackling Zero Overlaps}

We have assumed so far that 
$ \braket{A_N | B_N } = \det ( \mathbf{S} ) \neq 0 $. 
Thus, \Eq{\ref{eq: gen_wick}} is not applicable when $ \det ( \mathbf{S} ) = 0 $. 
This happens when \textbf{S} is rank-deficient.
We may also encounter cases where $ \det ( \mathbf{S} ) \approx 0 $, which can lead to numerical issues in \Eq{\ref{eq: rho_final}}. 
The near-zero scenario can be diagnosed by the SVD of the \textbf{S} matrix with an additional $\ComCom{N^3}$ cost. 

Recipes that tackle these zero or near-zero cases bypassing \Eq{\ref{eq: gen_wick}} exist, for example, as shown in \Reference{\citenum{RodriguezLaguna2020efficient}}.
We discuss the relevant results below.
The central object here is
$ \braket{A_N | \: \FA{q_d} \cdots \FA{q_1} \: \FC{p_1} \cdots \FC{p_d} \: | B_N } $, 
where we have reversed the normal ordering with respect to \Eq{\ref{eq: gen_wick}}.
This is because the transition matrix elements will be represented via an additional set of overlap matrix elements, 
as defined below 
\begin{subequations}
\begin{align}
S_{jk}^{(1)}
&= \braket{- | \: \FA{q_j} \: \FC{p_k} \: | -} 
= \delta_{q_j, p_k},
\\
S_{jk}^{(2)}
&= \braket{- | \: \FAlt{j} \: \FC{p_k} \: | -} 
= A_{p_k, j}^*,
\\
S_{jk}^{(3)}
&= \braket{- | \: \FA{q_j} \: \FCrt{k} \: | -} 
= B_{q_j, k}. 
\end{align}
\end{subequations}
Then it can be shown that \cite{RodriguezLaguna2020efficient}
\begin{align}
&\braket{A_N | \: \FA{q_d} \cdots \FA{q_1} \: \FC{p_1} \cdots \FC{p_d} \: | B_N }
\nonumber
\\
&= \det \begin{bmatrix}
\mathbf{S}_{d \times d}^{(1)} & \mathbf{S}_{d \times N}^{(3)} \\
\mathbf{S}_{N \times d}^{(2)} & \mathbf{S}_{N \times N}
\end{bmatrix},
\end{align}
where the \textbf{S} matrix elements are defined in \Eq{\ref{eq: overlap_mat_elems}}.
Applying the Schur complement formula leads to the following relation 
\begin{align} \label{eq: schur_form}
&\det \begin{bmatrix}
\mathbf{S}^{(1)} & \mathbf{S}^{(3)} \\
\mathbf{S}^{(2)} & \mathbf{S}
\end{bmatrix}
\nonumber
\\
&= \det ( \mathbf{S} ) \: \det \Big[ 
\mathbf{S}^{(1)} - \mathbf{S}^{(3)} \: \mathbf{S}^{-1} \: \mathbf{S}^{(2)} \Big].
\end{align}
The SVD of the overlap matrix \textbf{S} can be written as 
\begin{equation}
\mathbf{S} 
= \mathbf{U} \: \mathbf{\Sigma} \: \mathbf{V}^\dagger, \quad
\mathbf{\Sigma} 
= \begin{pmatrix}
\bm{\Sigma}^R & \mathbf{0} 
\\
\mathbf{0} & \bm{\Sigma}^E
\end{pmatrix}, 
\end{equation} 
where $ \bm{\Sigma}^R 
= \text{diag} ( \sigma_1, \dots, \sigma_{N - k}) $ 
and $ \bm{\Sigma}^E
= \text{diag} ( \epsilon_1, \dots, \epsilon_k ) $, with 
$ \{ \epsilon_j \} $ being a set of small numbers. 
Define 
$ ( \mathbf{U}_1 )_{N \times d}
= \mathbf{U}_{N \times N}^\dagger \: 
\mathbf{S}_{N \times d}^{(2)} $ and 
$ ( \mathbf{V}_1 )_{d \times N}
= \mathbf{S}_{d \times N}^{(3)} \: \mathbf{V}_{N \times N} $, 
followed by the partitioning
\begin{equation}
\mathbf{U}_1
= \begin{pmatrix}
\mathbf{U}_1^R \\
\mathbf{U}_1^E
\end{pmatrix}, \quad
\mathbf{V}_1
= \begin{pmatrix}
\mathbf{V}_1^R & \mathbf{V}_1^E
\end{pmatrix},
\end{equation}
where the superscripts $R$ and $E$ indicate the parts corresponding to $\bm{\Sigma}^R$ and $\bm{\Sigma}^E$, respectively. 
The $ \mathbf{U}_1^R $ and $ \mathbf{V}_1^R $ matrices have dimensions of $ (N - k) \times d $ and $ d \times (N - k) $, respectively, while the $ \mathbf{U}_1^E $ and $ \mathbf{V}_1^E $ matrices have dimensions of $ k \times d $ and $ d \times k $, respectively. 
Then it can be shown, starting from \Eq{\ref{eq: schur_form}}, that the transition matrix elements are \cite{RodriguezLaguna2020efficient} 
\begin{equation} \label{eq: trdm_modified}
\det \begin{pmatrix}
\mathbf{S}^{(1)} & \mathbf{S}^{(3)} \\
\mathbf{S}^{(2)} & \mathbf{S}
\end{pmatrix}
= e^{\mi \varphi} \det ( \bm{\Sigma}^R ) \: 
\det ( \bm{\mathcal{D}} )   \: 
\det ( \bm{\mathcal{C}} ),  
\end{equation}
where we have defined 
\begin{subequations} 
\begin{align}
e^{\mi \varphi}
&= \det ( \mathbf{U} ) \: \det ( \mathbf{V}^\dagger ),
\\
\bm{\mathcal{D}}_{k \times k}
&= - \mathbf{U}_1^E \: \bm{\mathcal{C}}^{- 1} \: 
\mathbf{V}_1^E,
\\
\bm{\mathcal{C}}_{d \times d}
&= \mathbf{S}^{(1)} 
- \mathbf{V}_1^R \: 
\big( \bm{\Sigma}^R \big)^{- 1} \: 
\mathbf{U}_1^R.
\end{align}
\end{subequations}
Thus, 
$ \braket{A_N | \: c_{q_d} \cdots c_{q_1} \: c_{p_1}^\dagger \cdots c_{p_d}^\dagger \: | B_N} $ is a definite scalar even when all $ \epsilon_j \rightarrow 0 $ in \Eq{\ref{eq: trdm_modified}}. 

A separate special case is when $\mathbf{S}$ becomes exactly a zero matrix. 
In this case, $\ket{A_N}$ and $\ket{B_N}$ are orthogonal, and differ by at least $N$ particle-hole excitations.
For example,
$ \ket{B_N} = \FClt{N + 1} \cdots \FClt{2N} \ket{-} $, where we have assumed $M \geq 2 \: N$.
Non-zero transition matrix elements are nevertheless possible if the operator $\mathbb{O}$ itself contains the full $N$-body particle-hole excitation structure 
required to map $\ket{B_N}$ into $\ket{A_N}$, which can be dealt with standard CAR algebra. 

\subsection{Summary}

The first step for computing $ \braket{A_N | \: \mathbb{O} \: | B_N} $, with $\mathbb{O}$ being any number-conserving fermionic operator, 
is to compute the overlap matrix \textbf{S}, which has $\ComCom{M N^2}$ cost. 
This is followed by a SVD of \textbf{S} with an $\ComCom{N^3}$ cost. 

If \textbf{S} is numerically full-rank, i.e., all singular values are above a numerical threshold, then 
the next step is to compute  $ \braket{A_N | B_N} = \det (\mathbf{S}) $ with an $\ComCom{N^3}$ cost. 
Then, any $ \braket{A_N | \: \mathbb{O} \: | B_N} $ can be computed in $\ComCom{1}$ time following \Eq{\ref{eq: gen_wick}}.

If \textbf{S} is rank-deficient, each $ \braket{A_N | \: \mathbb{O} \: | B_N} $ has to be handled separately by building
$ \mathbf{S}^{(2)}, \mathbf{S}^{(3)}, $ and the related matrices, followed by the computation in \Eq{\ref{eq: trdm_modified}}.
The dominant cost among all of these steps is $\ComCom{N^2 d}$, where we have assumed $\mathbb{O}$ is a $d$-body operator, which is usually a small integer. 


\section{Projection Steps After Numerical Integration} \label{app:projection}

Due to the multiple numerical integration steps, Hermiticity and idempotency of the 1-RDMs are not assured at every step of the propagation, 
and the $\bm{\gamma}_{n+1}$ obtained after RK4 in \Eq{\ref{eq:RK4step}}, more appropriately labeled as $ \bm{\gamma}^{\textrm{raw}}_{n+1}$, may no longer be a projector. 
To keep the single-SD ansatz at all times, we first impose Hermiticity, followed by an eigendecomposition
\begin{equation}
    \bm{\gamma}'_{n+1} 
    = \frac{1}{2} \: \Big[ \bm{\gamma}^{\textrm{raw}}_{n+1} 
    + \big( \bm{\gamma}^{\textrm{raw}}_{n+1} \big)^\dag \Big], \quad 
    \bm{\gamma}'_{n+1} 
    = \mathbf{U} \: \mathbf{\Lambda} \: \mathbf{U}^\dagger,
\end{equation}
then project to idempotency by replacing $\Lambda$ with a diagonal matrix where the first $N$ elements (occupied orbitals) are one, and the rest of the $ M - N $ elements (unoccupied orbitals) are zero,
\begin{equation}
\bm{\Theta}
= \text{diag} \big( \underbrace{1,\dots,1}_{N}, \underbrace{0,\dots,0}_{M-N} \big),
\end{equation}
so that the reconstructed 1-RDM is
\begin{equation}
\bm{\gamma}_{n+1}
= \mathbf{U} \: \bm{\Theta} \: \mathbf{U}^\dagger,
\end{equation}
where we assumed that we sorted eigenvectors based on eigenvalues if necessary. 
The corresponding coefficient matrix $\mathbf{C}_{n+1}$ for the new SD $|\psi(t_{n+1})\rangle$ is taken as the matrix of eigenvectors $\mathbf{U}^\dag$. 

However, it should be noted that since any unitary rotation within the occupied or unoccupied block leaves $\bm{\gamma}$ unchanged, additional alignment steps may be necessary to ensure smooth time-evolution of $\mathbf{C}(t)$. 
Explicitly, if $\mathbf{C}^{\mathrm{occ}}$ is the $M\times N$ block of occupied orbitals and $\mathbf{C}^{\mathrm{unocc}}$ the $M \times (M-N)$ block of unoccupied orbitals, then any unitary rotation applied to each block independently will produce the same 1-RDM elements. 
While this freedom is irrelevant for static properties, it can lead to unphysical features in dynamical processes. 
To maintain a smooth behavior, at each time step, we align the orbitals between $\mathbf{C}_{n}$ and $\mathbf{C}_{n+1}$ by solving the unitary Procrustes problem. \cite{Schonemann1966generalized, Schrader2023accelerated}
Focusing on the occupied subspace (with an identical procedure for the unoccupied subspace), one first computes the overlap matrix
\begin{equation}
   \mathbf{S}^{\mathrm{occ}} 
   = (\mathbf{C}^{\mathrm{occ}}_{n})^\dagger \: \mathbf{C}^{\mathrm{occ}}_{n+1},
\end{equation}
followed by a SVD of the $\mathbf{S}^{\mathrm{occ}}$ matrix,
\begin{equation}
    \mathbf{S}^{\mathrm{occ}} = \mathbf{U}^{\mathrm{occ}} \: \mathbf{\Sigma}^{\mathrm{occ}} \: \big( \mathbf{V}^{\mathrm{occ}} \big)^\dagger.
\end{equation}
Then, the rotation matrix that minimizes the distance between $\mathbf{C}^{\mathrm{occ}}_{n}$ and $\mathbf{C}^{\mathrm{occ}}_{n+1}$ is defined as
\begin{equation}
    \mathbf{R}^{\mathrm{occ}} 
    = \mathbf{V}^{\mathrm{occ}} \: \big( \mathbf{U}^{\mathrm{occ}} \big)^\dagger \: 
    \Rightarrow \: \tilde{\mathbf{C}}^{\mathrm{occ}}_{n+1} 
    = \mathbf{C}^{\mathrm{occ}}_{n+1} \: \mathbf{R}^{\mathrm{occ}}.
\end{equation}
Alternatively, the alignment procedure may be achieved using approaches based on Thouless rotations. \cite{Deumens1994, Tsereteli2006}


\bibliographystyle{apsrev4-2}
\bibliography{Ref}


\end{document}